\documentstyle[12pt]{article}

\def\fun#1#2{\lower3.6pt\vbox{\baselineskip0pt\lineskip.9pt
        \ialign{$\mathsurround=0pt#1\hfill##\hfil$\crcr#2\crcr\sim\crcr}}}

\renewcommand\({\left(}
\renewcommand\){\right)}
\renewcommand\[{\left[}
\renewcommand\]{\right]}

\newcommand\eq[1]{Eq.~(\ref{#1})}
\newcommand\eqs[2]{Eqs.~(\ref{#1}) and (\ref{#2})}

\newcommand\eqst[2]{Eqs.~(\ref{#1})--(\ref{#2})}

\newcommand\ee{\end{equation}}
\newcommand\be{\begin{equation}}
\newcommand\eea{\end{eqnarray}}
\newcommand\bea{\begin{eqnarray}}

\newcommand\sunit{\,\mbox{s}}

\newcommand\km{\,\mbox{km}}

\newcommand\GeV{\,\mbox{GeV}}
\newcommand\MeV{\,\mbox{MeV}}

\newcommand\Mpc{\,\mbox{Mpc}}

\newcommand\mpl{M_{\rm P}}

\newcommand\lsim{\mathrel{\rlap{\lower4pt\hbox{\hskip1pt$\sim$}}
    \raise1pt\hbox{$<$}}}
\newcommand\gsim{\mathrel{\rlap{\lower4pt\hbox{\hskip1pt$\sim$}}
    \raise1pt\hbox{$>$}}}

\newcommand\diff{\mbox d}

\def\dslash{\not{\hbox{\kern-2pt $\partial$}}}
\def\Dslash{\not{\hbox{\kern-4pt $D$}}}
\def\Oslash{\not{\hbox{\kern-4pt $O$}}}
\def\Qslash{\not{\hbox{\kern-4pt $Q$}}}
\def\pslash{\not{\hbox{\kern-2.3pt $p$}}}
\def\kslash{\not{\hbox{\kern-2.3pt $k$}}}
\def\qslash{\not{\hbox{\kern-2.3pt $q$}}}

 \newtoks\slashfraction
 \slashfraction={.13}
 \def\slash#1{\setbox0\hbox{$ #1 $}
 \setbox0\hbox to \the\slashfraction\wd0{\hss \box0}/\box0 }
 
\def\ee{\end{equation}}
\def\be{\begin{equation}}

\newcommand\sub[1]{_{\rm #1}}
\newcommand\su[1]{^{\rm #1}}

\begin{document}

\begin{flushright}
LANCS-TH/9919
\\hep-ph/9909387\\
(September 1999)
\end{flushright}
\begin{center}
{\Large \bf The abundance of moduli, modulini and gravitinos produced by
the vacuum fluctuation}

\vspace{.3in}
{\large\bf  David H.~Lyth}

\vspace{.4 cm}
{\em Department of Physics,\\
Lancaster University,\\
Lancaster LA1 4YB.~~~U.~K.}

\vspace{.4cm}
\end{center}

\begin{abstract}
Moduli, modulini and the gravitino have gravitational-strength interactions,
and thermal collisions after reheating 
create all of them with roughly the same 
abundance. With their mass of order $100\GeV$, corresponding to
gravity-mediated supersymmetry breaking, 
this leads to the well-known bound $\gamma T\sub R\lsim 10^9\GeV$
on the reheat temperature, where $\gamma\leq 1$ is the entropy
dilution factor. 
The vacuum fluctuation also creates these particles, 
with abundance determined by the solution of the 
equation for the mode function. Taking the equation in each case 
to be the one corresponding to a free field, we 
consider carefully the behaviour of the effective mass 
during the crucial era after inflation.
It may have a rapid oscillation, which does not however 
affect the particle abundance.
Existing estimates are confirmed; 
the abundance of modulini and (probably) of moduli created from 
the vacuum is less than from thermal collisions, but the abundance of 
gravitinos may be much bigger, 
leading to a tighter bound on $T\sub R$ if supersymmetry breaking is
gravity-mediated. 
\end{abstract}

\paragraph{Introduction} 
Long-lived particle species 
constrain models of the early Universe, because they
are cosmologically dangerous. In this note we are concerned with
species that have gravitational-strength interactions.
According to present ideas, these include the gravitino (spin $3/2$),
and species usually called moduli (spin $0$) and modulini (spin $1/2$).

The abundance of a given species 
is conveniently specified by the ratio $n/s$, where
$n$ is the number density and $s$ is the entropy density, 
evaluated at the epoch of nucleosynthesis.
A particle with a gravitational-strength decay rate, and
mass $100\MeV$ to $10^5\GeV$, decays before the present but after
nucleosynthesis. To avoid interfering with nucleosynthesis its
abundance must satisfy
\cite{subir}
\be
\frac n s \lsim 10^{-13}
\,.
\label{nscon}
\ee
(The precise bound lies between $10^{-12}$ and $10^{-15}$, depending on 
the mass.)
If supersymmetry breaking is gravity-mediated, the masses of the
gravitino, moduli and modulini are all expected to 
lie in this range. 

A particle with gravitational-strength interactions
and mass $\lsim 100\MeV$ is stable (lifetime longer than the age of the 
Universe). This is expected to be the case for
moduli, modulini and the gravitino, if supersymmetry
breaking is gauge-mediated.
Any stable species which is non-relativistic
at the present epoch has 
matter density 
\be
\Omega_X = 4.4 \times 10^9 h^{-2}
\frac{m_X}{1\GeV} \frac n s \lsim 0.3
\label{omcon}
\,,
\ee
where $h\simeq 0.65$ is the Hubble constant in units of $100\km\sunit^{-1}
\Mpc^{-1}$. 

Gravitinos are created by 
particle collisions after reheating, with abundance 
at nucleosynthesis \cite{subir}
\be
\(\frac ns\)\sub{thermal} 
\sim 10^{-13}
 \frac{\gamma T\sub{R}}{10^9\GeV}
\,.
\label{ntherm}
\ee
In this expression, $T\sub R$ is the reheat temperature,
and $\gamma^{-1}\geq1$ is the increase in entropy, per comoving volume,
between reheating and nucleosynthesis.
Moduli and modulini are
created with roughly the same abundance.
If \eq{nscon} holds for moduli, modulini or the gravitino,
we arrive at the famous bound $
\gamma T\sub{R}\lsim 10^9 \GeV$.

\paragraph{Creation from the vacuum}

The vacuum fluctuation during inflation provides another mechanism
for creating particles. Indeed, although there are no particles
during inflation, each field has  a vacuum fluctuation.
During some era containing the end of inflation,
the evolution of the mode functions may become non-adiabatic 
so that particles are created \cite{parker}.
After creation stops,
the number density is 
\be
n=n_* (a_*/a)^3
\,,
\ee
where $a$ is the scale factor of the Universe, and $a_*$ is its value at
the end of inflation. We shall refer to the constant $n_*$ as the number
density at the end of inflation, even though particle creation stops 
only some time later.

If the particles have only gravitational-strength interactions,
or are very heavy \cite{ckr,ckr23}, they will not 
thermalize at reheating. The particles created by the vacuum fluctuation
retain their identity, and 
\cite{lrs} their abundance at nucleosynthesis is
\bea
\frac ns &\sim& \(\frac{n_*}{V^\frac34}\) \(\frac{\gamma T\sub R}{V^\frac14} \)
\\
&\sim& 10^{-14} \( \frac{n_*}{H_*^3} \)
\(\frac{H_*}{10^{14}\GeV} \)
\(\frac{\gamma T\sub{R}}{10^9\GeV} \)
\,.
\label{nsest}
\eea
In this expression, $H_*$ is the practically constant value
of the Hubble parameter during inflation (to be precise, the value
at the end of inflation), and $V=3\mpl^2H^2$ is the inflationary 
potential. The observed cosmic microwave background 
anisotropy requires \cite{areview}
$H_*\lsim 10^{14}\GeV$ ($V^\frac14_*\lsim 10^{16}\GeV$). 
The corresponding number density,
for a stable non-relativistic species 
mass $m_X$, is
\bea
\Omega_X &\sim& 10^{10} \(\frac{n_*}{V_*^\frac34}\)
\(\frac{m_X}{1\GeV} \) \(\frac{T\sub R}{V_*^\frac14} \) \gamma \\
&\sim& 10^{10}\( \frac{n_*}{H_*^3} \)
\(\frac{H_*}{10^{14}\GeV} \)
\(\frac{m_X}{10^{14}\GeV}\)
\(\frac{T\sub{R}}{10^9\GeV} \) \gamma
\label{omest2}
\,.
\eea

If $n_*\lsim 10 H_*^3$, 
the thermal mechanism definitely
dominates \cite{lrs}, but if $n_*$ is bigger the vacuum fluctuation
may dominate, leading to a tighter constraint on $T\sub{R}$.
In terms of the occupation number
$|\beta_k|^2$ for 
states with momentum $k/a$, 
the number density per
helicity state of a given species is 
\be
n=\frac1{2\pi^2} \frac1{a^3} \int^{k\sub{max}}_0
|\beta_k|^2 k^2 \diff k
\,.
\label{nofbeta}
\ee
Significant particle creation
will occur only up to some $k\sub{max}$, such that
$|\beta_k|\sim 1$ at $k=k\sub{max}$, and $|\beta_k|\ll 1$ 
at bigger $k$. 

For fermions, $|\beta_k|\leq 1$ leading to the rather firm estimate
\cite{lrs}
\be
n_*\simeq 
10^{-2} (k\sub{max}/a_*)^3
\,.
\label{nest2}
\ee

For scalar particles, this expression gives in general only a lower bound
on $n_*$, because $|\beta_k|$ can
be much bigger than 1, but as we shall discuss it
is likely to give very roughly the right order of magnitude for moduli.

For a given particle species, $k\sub{max}$ is determined by 
the relevant mode function equation. If we are dealing with a free
field, this equation corresponds to 
the Klein-Gordon (spin 0) Dirac (spin $1/2$) or Rarita-Schwinger 
(spin $3/2$) equation, plus 
constraints to eliminate unwanted degrees of freedom
except in the case of scalars.
The resulting equations for spin zero and spin $1/2$
are well known, and they have
recently been given also for
helicity $3/2$ gravitinos \cite{mm,kklv,grt,l}, and 
for helicity $1/2$ gravitinos \cite{kklv,grt}. 

In the following we consider each case in turn, 
paying special attention to the behaviour of the effective
mass that appears in the mode function equation.
It may oscillate rapidly just after inflation, but this does not
affect the estimate of $k\sub{max}$.
In agreement with existing estimates, we find that
$k\sub{max}\sim a_* m_*$, where $m_*$ is the typical 
magnitude of the effective mass
soon after the end of inflation. Also in agreement with these 
estimates we find that for moduli and modulini,
$m_*\sim H_*$, while for helicity $1/2$ gravitinos
$m_*\sim M$ where $M\gsim H_*$ is the mass of the field responsible for
the inflaton potential.\footnote
{Throughout this paper, we assume that oscillation of the field
responsible for the inflaton potential is
dominated by its the mass term. This excludes the case
of inflation with a quartic potential. In that case,
the abundance of particles
created by the quantum fluctuation is enhanced, being the same as it
would be in the case of instant reheating \cite{kklv}.
In the case of the gravitino, we find that the advertised estimate
is reduced in the unlikely event that 
the gravitino mass (appearing in the Rarita-Schwinger equation)
is much bigger than $H$ around the end of inflation.}

\paragraph{Scalar particles}

The case of scalar particles, in particular moduli, is more 
complicated than the case of spin $1/2$ particles. The basic
reason is that the occupation number $|\beta_k|$ can be much
bigger than 1. As a result the number density \eq{nofbeta}
may be dominated by long wavelength modes. Related to this is the fact 
that the classical scalar field can have a practically homogeneous
value, which in the early Universe is different from the vacuum
value. In that case, the field eventually oscillates about the vacuum
value, corresponding to the existence of particles with negligible
momentum. 

For a given field $\phi$, a
crucial quantity is the effective mass-squared $m^2(t)$ in the early 
Universe. (We take the origin to be a fixed point of the symmetries
so that $V=V_0+\frac12m^2\phi^2$ plus higher terms.)
Assuming supergravity, it is typically
of order $\pm H^2$, until $H$ falls below the true mass $m_X$.
Consider first the value during inflation \cite{msquared,cllsw,ccqv}. 
Unless it is dominated by the $D$ term, the
potential $V$ is of the form
$\exp(K/\mpl^2)X$,
where $K$ is the K\"ahler potential, and the form of $X$ does not 
concern us for the moment. We are assuming canonical 
normalisation, which means that $K=\frac12\phi^2$ to leading order. As a result
the potential gives a contribution of the form
\be
m\sub{pot}^2 = V/\mpl^2 + \cdots 
\,.
\ee
The other contribution depends on $X$, and is of order $\pm
V/\mpl^2$. In general there is no reason to expect a cancellation,
so we expect
$m\sub{pot}^2=\lambda V/\mpl^2$,
with $\lambda\sim \pm1$. 
This is the only contribution to $m^2$ during inflation,
giving a practically constant value $m^2\sub{pot}=3\lambda H^2$.

Now consider the era just after inflation.
The field $\chi$ responsible
for $V$ oscillates, with angular frequency
$M$ equal to its mass. As a result $V$ and $m\sub{pot}^2$ 
oscillate with the same frequency.
In addition, there will be a contribution
\cite{drt,grt} 
$m^2\sub{kin}=\lambda' \dot\chi^2/\mpl^2$ with $\lambda'$ of order
$\pm1$ but {\em unrelated} to $\lambda$. 
During this era, the 
energy density $\rho=3H^2\mpl^2$ and the pressure $P$ are given by
\bea
\rho &=& V + \frac12\dot\chi^2 \label{rho}\\
P &=& -V + \frac12\dot\chi^2 \label{P}
\eea
The quantities $V$ and $\dot\chi^2$ oscillate 
out of phase, so that $H^2$ does not oscillate. In fact, 
energy conservation 
gives
$\dot H/H^2 = -\frac32(1+P/\rho)$, and according to field theory
$|P|\leq \rho$. If $\lambda$ and $\lambda'$ were equal, $m^2(t)$ would
have the same behaviour, but as they are not it oscillates. Redefining
the couplings we have
\be
m^2 \simeq H^2 \( \lambda + \lambda' \cos (Mt) \)
\,,
\ee
with $|\lambda|\sim|\lambda'|\sim 1$.

The energy in the oscillation of $\chi$ 
will sooner or later be
spread among several particle species, through reheating or earlier
during preheating (through parametric resonance). 
Adding together the incoherent 
equivalent field oscillations, we shall now get a 
non-oscillating contribution $m^2(t)\sim \pm H^2$.

Assuming that $m^2\sim \pm H^2$ the cosmology of a scalar field 
depends crucially on the sign. If it is negative,
the classical field value will be driven away from the fixed point,
and will end up oscillating about its
vev, corresponding to particle production that has nothing to do
with the vacuum fluctuation. It is this case that is
generally expected to occur for, at least some, moduli. 
In the case of gravity-mediated supersymmetry breaking, the estimated
abundance of moduli  is cosmologically dangerous. This is the usual 
version of the moduli problem \cite{modprob}.
If, on the other hand, $m^2$ is 
positive
during and after inflation, the classical value is always at the fixed
point of the symmetries, 
and particle creation from the vacuum fluctuation becomes relevant.
This case, corresponding to the 
field held at a fixed point of the symmetries, may occur for at
least some of the moduli as discussed for instance in \cite{glm}.
We are trying to see if the moduli problem occurs also in this case.

Before continuing with the case $m^2\sim H^2$, let us note that
during inflation, it may be that $|m^2|\ll H^2$.
(This is actually mandatory for the inflaton field \cite{treview},
which is not however expected to be a modulus.)
If that happens,  then (assuming minimal coupling to gravity) the very
long wavelength quantum fluctuation acts as a classical field
displacement, which again eventually oscillates giving a situation
essentially the same as the one in which $m^2\lsim -H^2$
\cite{glv,grt,fkl}.\footnote
{This conclusion does not assume that $m^2/H^2$ 
is small also after inflation, which can hardly occur
since each particle species present gives a separate contribution to 
it. 
The calculation of \cite{lrs}
reaching an apparently different conclusion did not take into
account the very long wavelength fluctuation.}

Now we briefly recall the formalism for the vacuum fluctuation
(see for instance \cite{lrs}).
A suitably chosen mode function 
obeys the equation
\be
u'' + \omega^2 u =0
\,,
\label{modeeq}
\ee
where
\bea
\omega^2 &\equiv& k^2 + (a\tilde m)^2  \\
\tilde m^2&\equiv& m^2 - \(1- 6\xi \) \frac{a''}{a^3} 
\,,
\eea
and $\xi=0$ for minimal coupling to gravity.
In these expressions the prime denotes differentiation with respect to 
conformal time $\eta$, 
where $\diff \eta = \diff t/a$. (An overdot will denote 
$\diff/\diff t$.)
Einstein gravity gives
\be
\frac{a''}{a^3} = \frac12 H^2(1- 3P/\rho)
\,.
\ee
During inflation, $a''/a^3= 2H^2$, and during matter domination
$a''/a^3=\frac12 H^2$.

The mode function takes on a simple form 
during any era in which $\omega$ satisfies the
{\em adiabatic condition}. 
This is the condition that $\omega$ be 
slowly varying on the conformal
timescale $\omega^{-1}$, or to be precise the conditions
\be
|\omega' |\ll \omega^2 \ ,\hspace{2em} |\omega''|\ll \omega^3  \,.
\label{ad1}
\ee
During any era in which the adiabatic condition is satisfied,
approximate solutions of \eq{modeeq}
are $u\su{adia}$
and $u\su{adia\;*}$, where 
\be
u\su{adia}\equiv 
\omega^{-\frac12} \exp(-i\int^\eta_{\eta_0}\omega \diff \eta)
\,,
\label{uflat1}
\ee
and $\eta_0$ is arbitrary. (Over a time interval $\ll H^{-1}$,
these solutions reduce to the flat spacetime solutions.)

The adiabatic condition is 
satisfied at early times during inflation, and
one adopts the initial condition $u=u\su{adia}$,
corresponding to the vacuum.
At late times the adiabatic condition is again satisfied and
\be
u=\alpha u\su{adia} + \beta u\su{adia\;*}
\,,
\ee
with $|\alpha|^2-|\beta|^2=1$. The occupation number is $|\beta|^2$,
and an explicit formula is
\be
4\omega |\beta|^2 = |u'|^2 +\omega^2|u|^2
-2\omega
\,.
\ee

There is negligible particle creation if
$|\beta|^2\ll 1$. A brief failure of the adiabatic condition 
need not lead to significant particle creation. 
At least for practical purposes, the criterion for
significant particle creation is
the failure of 
\be
|\overline{\omega'}| \ll \omega^2 \,,
\label{ad2}
\ee
where the average is over a conformal time interval $\omega^{-1}$.
We shall call this the {\em weak adiabatic condition}.
It says that 
the fractional change in $\omega$ is small, in a typical conformal
time interval $\omega^{-1}$. To be convinced of its relevance,
one can imagine the numerical integration of \eq{modeeq}. 
Alternatively,
one can note that \eq{modeeq} describes a two-dimensional harmonic
oscillator with a time-varying frequency;
the initial condition corresponds to circular motion,
and the final condition will also correspond to circular motion
if the weak adiabatic condition is satisfied.

We want to estimate $k\sub{max}$, defined 
as the wavenumber above which particle creation is negligible.
Equivalently, $k\sub{max}$ is the value of $k$, above which the
weak adiabatic condition \eq{ad2} is satisfied.
We assume $|\xi|\lsim 1$, $m^2\sim H^2$, and
$\tilde m^2 \sim H^2$. If we took $\tilde m^2/H^2$ to be constant,
there would be no need for the average in \eq{ad2}.
Since $aH$ peaks at the end of inflation, and varies on the inverse 
timescale $H$, this would give
$k\sub{max}\sim a_* H_*$ as advertised.
In fact, $\tilde m^2$ changes at the end of inflation
on the inverse timescale $M\gsim H$. If we ignored the average,
this would give $k\sub{max}\sim a_*\sqrt {H_*M}$.
Restoring it, we have $\overline {\omega' }= (H_*/M)\omega'$,
leading again to $k\sub{max}\sim a_* H_*$.

We still have to check that \eq{nofbeta} is dominated by the upper limit,
giving $n_*\sim 10^{-2} H_*^3$.
Three independent calculations \cite{dv,lrs,grt} verify
that this result is roughly correct if $\tilde m^2/H^2$
is fairly close to 1 during inflation.\footnote
{Reference \cite{dv} took $|m|\ll H_*$ and $\xi\sim 1$, and solved
the mode equation
assuming $a\propto 
\exp(H_* t)$ during inflation, and either $a\propto t^{\frac23}$ (matter 
domination) or $a\propto t^{\frac12}$ (radiation domination)
afterwards, with $u$ and $u'$ continuous at the transition.
Reference \cite{lrs} took $\tilde m =C_H H$ (with $C_H$ a 
constant of order 1) and $\xi\lsim 1$,
and solved the mode equation by the same matching 
procedure, finding
$n_*\sim 10^{-3.5}H_*^3$. Reference 
\cite{grt}
made the same assumptions about $m$ and $\xi$, but solved the mode
equation numerically assuming a quadratic inflation potential.
The first two calculations ignore the oscillation, after
inflation, of $m$ and $P/\rho$. The third includes
the latter oscillation but still ignores the former.}
On the other hand, $n_*$ is much bigger if $\tilde m^2/H_*^2$ 
is significantly below 1 \cite{grt,fkl}, due to the long wavelength effect
mentioned earlier.

The conclusion is that the abundance of moduli created from the vacuum
is less than the abundance from thermal collisions, if $m^2/H^2$ is 
fairly close to 1 during inflation, 
which (in magnitude)
is the `best guess' from supergravity with the
potential dominated by the $F$ term. On the other hand, the abundance of
moduli created from the vacuum is enhanced if $m^2/H^2$ is significantly
below 1 during inflation. Related to this is the fact that a negative
value $m^2\lsim -H^2$ (during or after inflation) generates 
moduli through the oscillation of the 
classical field, with abundance typically much bigger than the abundance 
from thermal collisions.

Before leaving the subject of scalar particle creation, we
should mention the case of a stable particle with true mass
$m_X\gsim 10^{12}\GeV$ \cite{ckr,ckr23}.
Such a particle is a dark matte candidate (Wimpzilla) which is never in 
thermal equilibrium, and whose abundance from 
thermal collisions may be strongly suppressed. If
$m_X$ accidentally happens to be of order $H_*$,
one again has $k\sub{max}\sim a_*H_*$, 
leading again to $n_*\sim 10^{-2}H_*^3$ if \eq{nofbeta} is dominated
by the large $k$ limit.
Using a variety of approaches to the solution of the mode
equation, it was found \cite{ckr} that $n_*\sim 10^{-5}H_*^3$
(taking $\xi=\frac16$) in rough agreement with this estimate.
For a suitable choice of $H_*$ one has (\eq{omest2})
a dark matter candidate. Alternatively the Wimpzillas can be created 
through (suppressed) thermal collisions, parametric resonance 
or from a first-order phase transition \cite{ckr23}.

We end this discussion of scalar particle production by mentioning the 
exceptional case $\xi=1/6$ with $m\to 0$. The adiabatic condition
is then satisfied at all times and there is no 
particle production, but there is no reason to expect this case to 
occur. The lack of particle production reflects the fact that
the field equation is invariant under conformal transformations
of the metric, making the expanding universe indistinguishable from
flat spacetime.

\paragraph{Spin $1/2$ particles}
The formalism for the creation of spin $1/2$ particles is described
for instance in \cite{lrs}. A suitably normalized two-component
mode function satisfies
\be
\( \begin{array}{c} u_+ \\ u_- \end{array} \)'
=i\( \begin{array}{cc} -am & k \\ k & am \end{array}  \)
\(\begin{array}{c} u_+ \\ u_- \end{array} \)
\,.
\label{half1}
\ee
The evolution is unitary so that one can choose
\be
|u_+|^2 + |u_-|^2 =1 
\label{unit}
\,.
\ee
Defining $u\equiv u_+$ one finds\footnote
{In \cite{lrs} $(ma)'$ is incorrectly replaced by $ma'$. 
With the ansatz $m/H=$constant that was used there, one can check
that 
this has no effect if inflation is followed by radiation domination,
and is equivalent
to the replacement $m/H\to \frac12 m/H$ if inflation is followed by 
radiation domination.}
\be
u'' + \(\omega^2 + i (ma)' \)u = 0 
\,,
\label{half2}
\ee
with
\be
\omega^2 = k^2 + (ma)^2
\,.
\label{half3}
\ee

If $am$ is slowly varying, 
solutions of the two-component mode equation are
\bea
u_\pm&=&u_\pm\su{adia}\equiv N_\pm
\exp\(-i\int^\eta_{\eta_0}\omega\diff \eta \)
\nonumber\\
u_\pm&=&v_\pm\su{adia}\equiv  N_\mp 
\exp\(i\int^\eta_{\eta_0}\omega\diff \eta \)
\label{adia}
\,,
\eea
where
\be
N_\pm =\pm \( \frac{\omega\pm am}{2\omega}\)^\frac12
\label{npm}
\,.
\ee
The ratio $N_+/N_-$ comes from \eq{half1}, and the 
normalization from \eq{unit}.
Over a time interval $\ll 
H^{-1}$, these are the usual
flat spacetime solutions. 
The adiabatic condition, ensuring that they 
are valid, is 
\be
(ma)' \ll \omega^2\ ,\hspace{4em} (ma)''\ll \omega^3
\,.
\label{adhalf1}
\ee
It may be obtained by inserting the solution into \eq{half2}
\cite{gprt}. Equivalently, one can note that
\eq{half1} requires $|N_\pm'/N_\pm|\ll \omega$, while the second 
differentiation to get \eq{half2} requires
also $|N_\pm''/N_\pm|\ll \omega^2$; the conditions on $N_-$ are the strongest, 
and lead to \eq{adhalf1}.

The adiabatic condition is satisfied at early times during inflation,
and at late times after inflation.
Taking $u_\pm=u_\pm\su{adia}$ at early times,
corresponding to the vacuum, one finds at late times
\be
u_\pm=\alpha u_\pm\su{adia} + \beta v_\pm\su{adia}
\,,
\label{half4}
\ee
with $|\alpha|^2+|\beta|^2=1$. 
This corresponds to
\be
u=\alpha N_+\exp\(-i\int^\eta_{\eta_0}\omega\diff \eta \)
+\beta N_- \exp\(i\int^\eta_{\eta_0}\omega\diff \eta \)
\,.
\ee
An explicit formula for the occupation 
number $|\beta|^2$ is
\be
|\beta|^2=\(\omega+am-2{\rm Im}(uu{^*}')\)/\(2\omega\)
\,.
\label{half5}
\ee

Analogously with the scalar case, a brief failure of the adiabatic 
condition will not cause significant particle creation, the condition 
for that being
\be
|\overline{(ma)' } | \ll \omega^2
\,,
\label{adhalf2}
\ee
where the average is over a conformal time interval $\omega^{-1}$.

The adiabatic condition is satisfied in the 
limit $m\to 0$, reflecting 
the invariance of the Dirac equation under conformal 
transformations of the metric. There is no particle creation in this 
limit.

In contrast with the case for scalars, spin
$1/2$ particles with gauge interactions
usually have effective mass $|m|\ll H$
in the early Universe \cite{ccqv}. There is essentially
no creation of such particles from the vacuum.

For gauge singlets, including 
modulini, the situation is similar to that which we discussed already
for scalars and spin $1/2$ particles. Supergravity is 
expected to generate an effective mass
$m\sim \pm H$,
coming from several different sources just as for scalar masses
\cite{ccqv}. The ratio $m/H$ is constant during inflation, 
and oscillates afterwards with angular frequency $M\gsim H_*$, before settling
down to a new constant value. 
Applying the weak adiabatic condition
\eq{adhalf2} (not the full condition \eq{adhalf1}, which would give
too big a value\footnote
{In a different context \cite{gprt}, $k\sub{max}$ 
is estimated on the assumption that there 
is significant particle production whenever the full adiabatic condition
\eq{adhalf1} is violated. We have checked that, in that case,
the same result would have been
obtained using the more appropriate weak adiabatic condition
\eq{adhalf2}.}
for $k\sub{max}$) this leads to the advertised
estimate $k\sub{max}\sim a_* H_*$, and $n_*\sim 10^{-2}H_*^3$,
leading to the conclusion that the creation of modulini from the
vacuum is less efficient than the thermal creation. 

Note that, in contrast with the scalar case, the abundance of 
spin $1/2$ particles is not altered
if $m/H$ happens to be very small 
during inflation, which may happen in some models.

\paragraph{The gravitino mode equations}

At the time of writing, the study of gravitino production is just 
beginning \cite{mm,kklv,grt,l}. One has to consider separately
the production of helicity $1/2$ and helicity $3/2$ 
particles, as seen by a comoving observer. 
If the gravitino in the early Universe can be treated as a free field,
it satisfies the Rarita-Schwinger equation, with 
time-dependent 
effective mass\footnote
{This is the well-known expression of $N=1$ supergravity.
In the early Universe, it may be reasonable to consider instead
\cite{juan} $N>1$.}
\be
m=e^{\frac12K/\mpl^2} |W|/\mpl^2
\,,
\label{gmass}
\ee
where $W$ is the superpotential (a holomorphic function of the 
complex scalar fields $\phi_n$) and 
$K$ is the K\"ahler potential (a real function of $\phi_n$
and $\phi_n^*$). In minimal supergravity, $K=\sum|\phi_n|^2$
but one expects additional terms suppressed by powers of $\mpl$.
The mode equations are found by 
imposing constraints to eliminate unwanted degrees of freedom.
For helicity $3/2$, one finds \cite{mm,kklv,grt,l} the same
equation, \eq{half1}, as in the spin $1/2$ case. 

The mode equation for helicity $1/2$ gravitinos
is more complicated \cite{kklv,grt}.
A suitably defined two-component mode function satisfies
\cite{grt}
\be
\( \begin{array}{c} u_+ \\ u_- \end{array} \)'
=i\( \begin{array}{cc} -am & kG \\ kG^* & am \end{array} \)
\(\begin{array}{c} u_+ \\ u_- \end{array} \)
\,.
\label{g1}
\ee
The function $G$ may be written
\be
G=e^{-i\theta} A^*
\,,
\label{g1a}
\ee
where 
$A=A_1+iA_2$ is the
quantity defined in \cite{kklv},
\bea
A_1 &\equiv& \frac{(P/\rho) H^2-m^2}{H^2+m^2} \nonumber\\
A_2 &\equiv & \frac23\frac{\dot m}{H^2+m^2}
\label{g2}
\,,
\eea
and 
\bea
\cos\theta&\equiv& -\frac{m^2-H^2}{m^2+H^2} \nonumber\\
\sin\theta&\equiv& \frac{2mH}{m^2+H^2} 
\,.
\label{g3}
\eea

The evolution  defined by \eq{g1} is unitary, leading to
\eq{unit}. The function $u=G^{-\frac12} u_+$, satisfies
\be
u'' + \(\omega^2 + i (\tilde m a)' \) u = 0
\label{g4}
\,,
\ee
with
\be
\omega^2 
\equiv |A|^2k^2 + (\tilde m a)^2 
\,,
\label{g5}
\ee
and 
\bea
\tilde m &=&  -\frac12 m - \frac32 m A_1- \frac32 H A_2 
-\frac1{2i} \frac{\dot A}{A} \label{g6}
\\
&=& m +\frac1{2i} \frac{\dot G}{G} 
\label{g7} \\
&\equiv& m + \frac12 (\arg G)\;\dot{} + \frac12
\frac{|G|\;\dot{}}{|G|}
\eea
The first expression is given in \cite{kklv}, while the second is
given in \cite{grt}; they are equivalent by virtue of
\eq{g1a}.\footnote
{To be more precise, the last terms of \eqs{g4}{g6} appear with opposite
sign in \cite{kklv}. The origin of this discrepancy, which may be partly
due to different conventions, is not clear and will not affect 
order-of-magnitude estimates of the abundance. References
\cite{kklv,grt} give \eqst{g4}{g7} only for the special case $|A|=1$, which
may be the only one of physical interest, but we have checked that
they follow from \eq{g1} also in the case $|A|\neq 1$.}

\paragraph{The gravitino abundance and its cosmological consequences}
The formalism of the last section applies if the 
gravitino can be treated as a free field in the early Universe,
corresponding to the Rarita-Schwinger equation. This has been demonstrated
\cite{kklv} only when the energy density and pressure are dominated
by the oscillation of a single homogeneous scalar field, with
minimal kinetic term and a 
potential (taken to be tree-level) dominated by the 
$F$  term of $N=1$ supergravity.
It has been shown \cite{kklv,grt} that in this case, $|A|=1$,
so that the production of helicity $1/2$ gravitinos is exactly the same
as production of a spin $1/2$ particle with mass $\tilde m$.
At least if the magnitude of the scalar field 
is small on the Planck scale,
this particle can be identified with the goldstino of global supersymmetry
\cite{kklv,grt}.

In the case $|A|=1$, the last term of \eq{g6} may be written
\be
\frac1{2i} \frac{\dot A}{A} = -\frac12\frac{\dot A_1}{A_2}
=\frac 32 m (1+ A_1)
-\frac{\dot w - 3H(1+w)(w-A_1)}{\[1-w^2 + 2\(1+w\) m^2/H^2 \]^\frac12
}
\,,
\label{tilm2}
\ee
where $w\equiv P/\rho$.
During slow-roll inflation, 
$w$ is close to $-1$ with $|\dot w|\ll H(1+w)^\frac12$,
and afterwards 
$w$ will oscillate between $\pm 1$, with angular  frequency 
equal to the mass $M\gsim H_*$ of the field 
responsible for the inflaton potential.
Bearing all this in mind, \eq{tilm2} shows that if
$m\lsim H$, $\tilde m$ increases sharply just after inflation, 
to a value of order $M$, in a time of order $M^{-1}$.
Since we are dealing with the 
spin $1/2$ mode equation, the adiabaticity condition
is $\overline{(a\tilde m)'}\ll \omega^2$,
leading to the estimate
$k\sub{max}\sim a_* M$ and $n_*\sim 10^{-2}M^3$.

This is the result already anticipated \cite{kklv,grt}.
It is indeed reasonable to suppose that $m\lsim H$, because 
according to supergravity $H^2=\frac13\rho(t)/\mpl^2$
always receives a contribution $-m^2(t)$.
At the present epoch this contribution is cancelled to high
accuracy for some unknown reason (part of
the cosmological constant problem)
but during inflation there is no reason to expect that to
happen, in the usual case that $H_*$ is bigger than the 
present value of the gravitino mass.

This estimate of the gravitino abundance is based on 
\eq{g4} for the mode function, taking $|A|=1$ so that
the equation becomes the same as in the spin $1/2$ case.
The equation $|A|=1$ (with $A$ defined by \eq{g2}) cannot
be exact, because it relates the gravitino mass $m(t)$,
 given by \eq{gmass},
to the density and pressure. After reheating, or significant
preheating, the density
and pressure have nothing to do with the scalar fields
that define $m$. One can even construct a model of inflation,
albeit an unrealistic one, in which $m(t)$ can be chosen independently 
of the density and pressure throughout the history of the Universe;
this is the model \cite{casas} which generates the inflaton potential
entirely from a $D$ term.

It may be that whenever $|A|$ is 
different from 1, \eqs{g1}{g4} cease
to be true.
Let us nevertheless discuss briefly the opposite possibility,
that the gravitino in the early Universe is described by the
Rarita-Schwinger equation (plus constraints)
with $|A|\neq 1$. This equation has well-known problems in a generic 
spacetime \cite{robin}, but it is not clear that they exist
in the special case of the Robertson-Walker Universe.
The phase velocity $|A|$
associated with \eq{g4} will be less than 1, as is presumably
required by causality, provided that $m$ does not vary too
rapidly. Coming to the question of gravitino creation,
the following argument suggests that the
estimate $k\sub{max}\sim a_* M$ is likely to survive.
Assuming as always $m\lsim H$, it is clear that $|\tilde m|a$ will 
have the same general behaviour as in the $|A|=1$ case, being 
of order $a_* M$ just after inflation and smaller at earlier
and later epochs. Also, 
$|A|$ is at least roughly of order 1 just after inflation,
since $P/\rho$ oscillates between $\pm 1$.
Choosing $k>a_* M$,
the mode equation therefore reduces to
$u''+\omega^2 u=0$, with $\omega=|A|k$. 
This is the same as \eq{modeeq}, and if \eq{ad2} is satisfied
the positive-frequency mode is never generated, and 
there is no particle creation. 
\eq{ad2} is just
\be
k > \overline{|A|'}/|A|^2
\,.
\ee
It follows that $k\sub{max}$ is of order $a_* M$ or the maximum of
$\overline{|A|'}/|A|^2$, whichever is bigger. By inspection of 
\eq{g2}, one can see that the latter quantity 
is typically of order $a_*M$ just after inflation and 
smaller at earlier and later times, so that indeed
$k\sub{max}\sim a_* M$.

Finally, let us look at the cosmological significance of the estimate
$n_*\sim 10^{-2} M^3$.
Depending on the model of inflation, $M$ can be anywhere 
between $H_*$ and (roughly) $V^{1/4}$, where $V^{1/4}\simeq (\mpl H_*)
^\frac12$ is the inflaton potential. 
In the extreme case $M\sim V^\frac14$,
\eq{nsest} gives \cite{grt}
\be
\frac ns \sim 10^{-2} \frac{T\sub R}{V^\frac14} \gamma \,.
\ee
In the case of gravity-mediated supersymmetry breaking,
$n/s\lsim 10^{-13}$ leading to a tight bound on $T\sub R$,
which becomes {\em more} stringent as the scale of inflation
decreases, with the nucleosynthesis bound $T\sub R\gsim
10\MeV$ imposing the constraint
\be
V^\frac14 > 10^9\GeV
\label{vlb}
\,.
\ee

\paragraph{Conclusion}

We have considered the abundance
of moduli, modulini and gravitinos created from the 
vacuum fluctuation, 
paying special attention to the behaviour of the
effective mass just after inflation.
In agreement with previous estimates, we find that 
creation from the vacuum is likely to be insignificant
for moduli and modulini, but may be very important for gravitinos.
These estimates are based on the assumption
that in the early Universe, the relevant equation
is the one corresponding to a free field.

\section*{Acknowledgements}
I thank Robin Tucker and Vladimir Falko
for a useful conversation, and thank the 
organisers of COSMO99 at ICTP, Trieste, where I had useful
conversations with Lev Kofman, Andrei Linde and Toni Riotto.

\newcommand\pl[3]{Phys. Lett. #1 (19#3) #2}
\newcommand\np[3]{Nucl. Phys. #1 (19#3) #2}
\newcommand\pr[3]{Phys. Rep. #1 (19#3) #2}
\newcommand\prl[3]{Phys. Rev. Lett. #1 (19#3) #2}
\newcommand\prd[3]{Phys. Rev. D #1 (19#3) #2}
\newcommand\ptp[3]{Prog. Theor. Phys. #1 (19#3) #2}
\newcommand\rpp[3]{Rep. on Prog. in Phys. #1 (19#3) #2}
\newcommand\jhep[2]{JHEP #1 (19#2)}
\newcommand\grg[3]{Gen. Rel. Grav. #1 (19#3) #2}

\end{document}